# Similarity-and-Independence-Aware Beamformer: Method for Target Source Extraction using Magnitude Spectrogram as Reference

*Atsuo Hiroe*[1]

[1]Sony Corporation
atsuo.hiroe@sony.com

## Abstract

This study presents a novel method for source extraction, referred to as the similarity-and-independence-aware beamformer (SIBF). The SIBF extracts the target signal using a rough magnitude spectrogram as the reference signal. The advantage of the SIBF is that it can obtain an accurate target signal, compared to the spectrogram generated by target-enhancing methods such as the speech enhancement based on deep neural networks (DNNs). For the extraction, we extend the framework of the deflationary independent component analysis, by considering the similarity between the reference and extracted target, as well as the mutual independence of all potential sources. To solve the extraction problem by maximum-likelihood estimation, we introduce two source model types that can reflect the similarity. The experimental results from the CHiME3 dataset show that the target signal extracted by the SIBF is more accurate than the reference signal generated by the DNN.

## 1. Introduction

Processes of extracting the target signal from mixtures of multiple sources, such as denoising and speech extraction, play a significant role in improving speech recognition performance [1]. Generally, the associated methods are classified into nonlinear and linear. In the last decade, the nonlinear methods have drastically improved owing to the development of deep neural networks (DNNs). These methods, referred to as DNN-based speech enhancements (SEs), can generate clean speeches from noisy ones [2][3] and extract an utterance from overlapping speeches [4][5]. However, linear methods, such as the beamformer (BF), are advantageous in the following aspects:

1. Avoiding nonlinear distortions, such as musical noises and spectral distortions [6][7].
2. Improving the quality of the extracted sound by increasing the number of microphones [8][9].
3. Estimating proper phases and scales of the extracted sound using designated techniques, such as rescaling in the independent component analysis (ICA) [10][11].

To incorporate the features of both the linear and nonlinear methods, we develop a new BF that uses the signal generated by any target-enhancing method (including DNN-based SE) as a reference signal. Since the magnitude spectrogram is more accessible than the time-frequency mask and complex-valued spectrogram, this study uses it as the reference. It should be noted that such a reference is considered "rough" (or less accurate) in the following regards:

a) It still includes some nondominant interferences (signals other than the target one), or can be distorted by the side effects of removing these interferences.
b) It does not contain any phase information.

The purpose of proposing the new BF is to generate an extracted target that is more accurate than the reference. The existing reference-based or DNN-based approaches, however, fail to meet this purpose, as discussed in Section **2**. Therefore, a novel method, the similarity-and-independence-aware beamformer (SIBF), is proposed, and presented in Section **3**.

## 2. Related works

The concept behind our proposed method is similar to that of the ICA with references (ICA-R) [12][13][14][15]. Particularly, the one-unit ICA-R [13] can generate a single signal corresponding to the reference. These approaches, however, do not consider the combination of the real-valued reference and complex-valued signals.

The independent deeply learned matrix analysis (IDLMA) is developed as a framework of the semiblind source separation [16]. In the IDLMA, the power spectrogram of each source is first estimated by DNNs; each source is then estimated using the power spectrogram as the reference. However, the IDLMA requires multiple references for all sources, including cases that only one source is of interest.

Other related works include combining the DNN for time-frequency mask estimation with the existing BFs, such as the minimum variance distortionless response (MVDR) BF

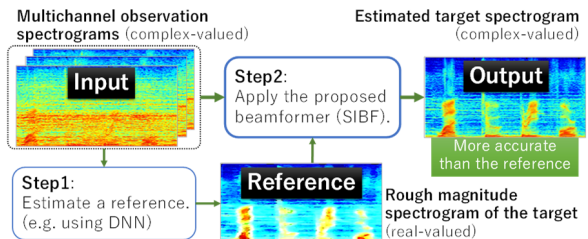

Figure 1: *Workflow of the proposed beamformer.*

Table 1: *Signal notations*

| Signal name | Spectrogram | Element | Column vector of all channel elements |
|---|---|---|---|
| **Source** | $S_k$ | $s_k(f,t)$ | $s(f,t)$ |
| **Observation** | $X_k$ | $x_k(f,t)$ | $x(f,t)$ |
| **Uncorrelated observation** | $U_k$ | $u_k(f,t)$ | $u(f,t)$ |
| **Estimated source** | $Y_k$ | $y_k(f,t)$ | $y(f,t)$ |
| **Reference** | $R$ | $r(f,t)$ | (not available) |

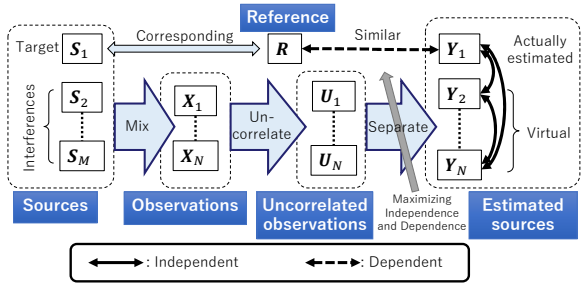

Figure 2: *Modeling procedure of the mixing and separating processes. The unique points of the modeling are (1) the dependence between $Y_1$ and $R$ is considered, as well as the independence of all the estimated sources, and (2) only $Y_1$ is actually estimated while $Y_2, \ldots, Y_N$ are just virtual sources.*

[17][18] and generalized eigenvalue (GEV) BF [19][20]. However, these methods do not achieve our purpose because they cannot directly treat the magnitude spectrogram.

## 3. Problem formulation of the similarity-and-independence-aware beamformer

The notations listed in Table 1 are used consistently throughout this paper to represent the time-frequency domain signals, with $f$, $t$, and $k$ denoting the indices of the frequency bin, frame, and channel, respectively.

The workflow of the proposed SIBF is shown in Figure 1. The inputs are the multichannel observation spectrograms obtained from multiple microphones, and the output is an extracted target spectrogram. A rough magnitude spectrogram of the target, which can be estimated using various methods including the DNN-based SE, is used as the reference.

The workflow involves two steps: (i) estimating the rough magnitude spectrogram of the target, and (ii) applying the SIBF with the rough spectrogram as the reference.

To realize this process, we extend the framework of the deflationary ICA using uncorrelations (prewhitening) [21], as presented in the subsequent subsections.

### 3.1. Mixing and separating processes with the reference

Figure 2 shows the modeling procedure of the mixing and separating processes of the SIBF. We assume that the sources $S_1, \ldots, S_M$ are mutually independent. Without loss of generality, $S_1$ is considered the target in this study. The observations $X_1, \ldots, X_N$ represent the spectrograms obtained from $N$ microphones. In the time–frequency domain, $X_k$ is approximated as the instantaneous mixture of the sources. We generate the uncorrelated observations $U_1, \ldots, U_N$ to apply the framework of the deflationary ICA. From the assumption of the sources, the estimated sources $Y_1, \ldots, Y_N$, which are the results of the separation, are also mutually independent.

The uncorrelation process in each frequency bin is expressed as (1), using the uncorrelation matrix $P(f)$:

$$u(f,t) = P(f)x(f,t) \text{ s.t. } \langle u(f,t)u(f,t)^H\rangle_t = I, \quad (1)$$

where $\langle \cdot \rangle_t$ and $I$ denote the averages over $t$ and the identity matrix, respectively. Similarly, the separating process is expressed by (2), using the separation matrix $W(f)$ to make $y_1(f,t), \ldots, y_N(f,t)$ mutually independent.

$$y(f,t) = W(f)u(f,t) = W(f)P(f)x(f,t). \quad (2)$$

Considering the uncorrelation, we can restrict $W(f)$ to a unitary matrix such that $W(f)W(f)^H = I$ [21].

To extract only the estimated target $y_1(f,t)$, we also use

$$y_1(f,t) = w_1(f)u(f,t) = w_1(f)P(f)x(f,t), \quad (3)$$

where $w_1(f)$ is the first row vector in $W(f)$.

The rest of Figure 2 shows the unique points of our modeling procedure. The reference $R$ is a rough estimate of the target $S_1$. To associate $Y_1$ with $S_1$, we consider the dependence between $Y_1$ and $R$ as well as the independence of the estimated sources. Conversely, maximizing the independence makes $Y_1$ more accurate than the reference, whereas maximizing the dependence only makes $Y_1$ similar to the reference.

Because $S_1$ is the only source of interest, we employ the deflationary estimation [21], i.e., one-by-one separation, to generate $Y_1$ only. This indicates that the other estimated sources, $Y_2, \ldots, Y_N$, are virtual (potential).

### 3.2. Maximum-likelihood estimation of the target signal

We solve the target extraction problem shown in Figure 2 by maximum-likelihood (ML) estimation, which is widely used in blind source separation (BSS) problems [16][22][23].

For the dependence between the estimated target and reference, we consider the temporally averaged negative log-likelihood (TANLL) of both observations and reference using

$$\text{TANLL} = -\frac{1}{T}\log p_{RX}(R, X_1, \cdots, X_N), \quad (4)$$

where $T$ is the number of frames and $p_{RX}$ denotes the joint probability density function (PDF) of its arguments. For simplicity, we assume that $M = N$ and all elements in the same spectrogram are mutually independent. From these assumptions and (2), we can rewrite (4), to obtain (5) and (6).

$$(4) = -\sum_f \langle \log p_{rx}(r(f,t), x(f,t))\rangle_t \quad (5)$$

$$= -\sum_f \langle \log p_{rs_1}(r(f,t), y_1(f,t))\rangle_t$$
$$\quad - \sum_f \sum_{k\geq 2} \langle \log p_{s_k}(y_k(f,t))\rangle_t$$
$$\quad - 2\sum_f \log|\det(W(f)P(f))|, \quad (6)$$

where $p_{rx}$, $p_{rs_1}$, and $p_{s_k}$ denote the joint PDF between $r(f,t)$ and $x(f,t)$, joint PDF between $r(f,t)$ and $s_1(f,t)$, and PDF of $s_k(f,t)$, respectively. We refer to $p_{rs_1}$ as a *source model*, which is examined in Section **3.3**.

By minimizing (6), we can estimate the most likely sources. Because of the uncorrelation, the determinant of $W(f)P(f)$ in (6) is constant. Therefore, to estimate $w_1(f)$, which is the extraction filter for $y_1(f,t)$, we minimize only the first term in (6), subject to $w_1(f)w_1(f)^H = 1$ as follows:

$$w_1(f) = \arg\min_{w_1(f)}\{-\langle \log p_{rs_1}(r(f,t), y_1(f,t))\rangle_t\} \quad (7)$$

### 3.3. Source models

To reflect the dependence onto the source model, we examine two types of PDFs: (i) the time-frequency-varying variance (TV) model, and (ii) bivariate spherical (BS) model.

The TV model includes different variances in each frequency bin and frame. The reference $r(f,t)$ is interpreted as a value related to the variance. This study then uses the TV Gaussian model, which has widely been used in BSS problems

[16][23][24]. To control the influence of the reference, we append $\beta$ as the *reference exponent*:

$$p_{rs_1}(r(f,t), y_1(f,t)) \propto \frac{1}{r(f,t)^{\beta/2}} \exp\left(-\frac{|y_1(f,t)|^2}{r(f,t)^\beta}\right). \quad (8)$$

In contrast, the BS model is a two-variable version of the multivariate spherical (MS) distribution [25]. MS models, such as the MS Laplacian model, are used in independent vector analysis (IVA) to avoid the permutation ambiguity problem because they can make all the frequency components dependent [25][26][27][28]. To make $|y_1(f,t)|$ and $r(f,t)$ mutually dependent, we use the following BS Laplacian model:

$$p_{rs_1}(r(f,t), y_1(f,t)) \propto \exp\left(-\sqrt{\alpha r(f,t)^2 + |y_1(f,t)|^2}\right), \quad (9)$$

where $\alpha$, called the *reference weight*, controls the influence of the reference. To balance the scales of $r(f,t)$ and $y_1(f,t)$, we normalize the reference such that $\langle r(f,t)^2 \rangle_t = 1$.

### 3.4. Rules for estimating the extraction filter

In this subsection, we derive the rules for estimating the filter for each model.

For the TV Gaussian model, the closed-form solution is written as (10), because assigning (8) to (7) results in the problem of minimizing a weighted covariance matrix.

$$\mathbf{w}_1(f) = \text{EIG}(\langle \mathbf{u}(f,t)\mathbf{u}(f,t)^H / r(f,t)^\beta \rangle_t)^H, \quad (10)$$

where $\text{EIG}(\cdot)$ denotes the eigenvector in the row vector form corresponding to the minimum eigenvalue of the given matrix.

For the BS Laplacian model, we apply iterative updating rules written as (11) and (12), which are based on the auxiliary function algorithm [29].

$$b(f,t) \leftarrow \sqrt{\alpha r(f,t)^2 + |\mathbf{w}_1(f)\mathbf{u}(f,t)|^2} \quad (11)$$

$$\mathbf{w}_1(f) \leftarrow \text{EIG}(\langle \mathbf{u}(f,t)\mathbf{u}(f,t)^H / b(f,t) \rangle_t)^H \quad (12)$$

To derive these rules, we use the following inequation, which contains a positive value $b(f,t)$ called the *auxiliary variable*:

$$\sqrt{\alpha r(f,t)^2 + |y_1(f,t)|^2} \leq \frac{\alpha r(f,t)^2 + |y_1(f,t)|^2}{2b(f,t)} + \frac{b(f,t)}{2}. \quad (13)$$

In the first iteration, we use $b(f,t) \leftarrow r(f,t)$ instead of (11) because $\mathbf{w}_1(f)$ is unknown. This is equivalent to the rule of the TV Gaussian model (10) with $\beta = 1$.

### 3.5. Rescaling in postprocess

After the filter estimation, we estimate the proper scale and phase of the estimated target using (3) and (14), which represent mapping to the specified observation signal and are equivalent to the projection-back or minimal distortion principle method [10][11].

$$y_1(f,t) \leftarrow \frac{\langle x_m(f,t)\overline{y_1(f,t)} \rangle_t}{\langle |y_1(f,t)|^2 \rangle_t} y_1(f,t), \quad (14)$$

where $m$ denotes the reference microphone index for rescaling.

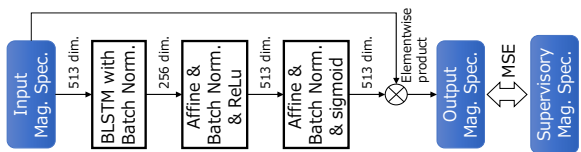

Figure 3: *Network configuration for DNN that outputs a magnitude spectrogram (Mag. Spec.) as the reference (numbers indicate the input and output dimensions.)*

## 4. Experiments

To verify the effectiveness of the proposed SIBF, we conducted several experiments using the CHiME3 dataset [30]. This means that we applied the SIBF to the problem of estimating clean speeches in noisy environments. In the dataset, sound data were recorded in four noisy environments using six microphones attached to a tablet device. Clean speeches were also recorded in a recording booth.

### 4.1. DNN for reference estimation

To prepare the DNN for reference estimation, we modified the configuration that trains the bidirectional long short-term memory (BLSTM) based mask estimator for GEV BF [19] to output the magnitude spectrogram. The network configuration, as shown in Figure 3, was similar to that in [19], except for the following aspects:

1) Supervisory data consisted of magnitude spectrograms of clean speeches instead of ideal binary masks.
2) The mean squared error (MSE) was used as the loss function in the training stage.
3) The training was performed in 20 epochs.

To estimate the reference, the observation spectrogram of Microphone #5 (closest to the speaker position) was used as the DNN input.

### 4.2. Experimental setups

To prepare the input data as a development set with various signal-to-noise ratios (SNRs), we artificially mixed the clean speeches recorded in the booth (BTH) with background (BG) noises. During the mixing, we applied four multipliers (0.25, 0.5, 1.0, and 2.0) to the noises, and termed the mixed data as "BTH+BG×0.25" (or simply "BG×0.25"), and so on.

In the preprocess, we converted the waveforms into spectrograms using the short-time Fourier transform with 1024 points and 256 shifts. In the postprocess, we employed (14) with $m = 5$, which was the index of the closest microphone.

### 4.3. Best parameter sets for each source model

To determine the best parameters for each model, we conducted a series of experiments using the perceptual evaluation of speech quality (PESQ) as the performance score.

First, we evaluated the TV Gaussian model (8) to find the best reference exponent $\beta$. For cases $\beta = 0.5, 1, 2, 4,$ and $8$, we found that the $\beta = 8$ case achieved the best PESQ score, while the $\beta = 2$ case demonstrated the worst, although only the latter case strictly corresponds to the TV Gaussian distribution.

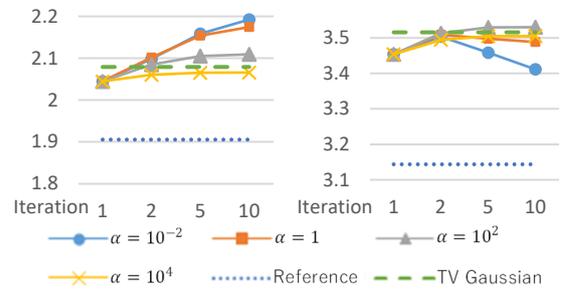

Figure 4: *PESQ scores of BS Laplacian model for reference weight $\alpha$, reference, and TV Gaussian model ($\beta = 8$) (left: BTH+BG×2.0 scenario (lower SNR); and right: BTH+BG×0.25 scenario (higher SNR))*

Table 2: *PESQ and SDR for all methods; NN-SIBF: SIBF using the reference generated by DNN; Oracle SIBF: SIBF using ideal references; "× 0.25", ..., "× 2.0": multiplier of noise in mixing speeches and noise; and Eval: CHiME3-simulated evaluation set (best score for each scenario is bolded (Oracle SIBF is not considered in the comparison))*

| Method | Source Model | PESQ | | | | | SDR [dB] | | | | |
|---|---|---|---|---|---|---|---|---|---|---|---|
| | | $\times 0.25$ | $\times 0.5$ | $\times 1.0$ | $\times 2.0$ | Eval | $\times 0.25$ | $\times 0.5$ | $\times 1.0$ | $\times 2.0$ | Eval |
| NN-SIBF | TV Gaussian | 3.52 | 3.12 | 2.63 | 2.08 | 2.67 | 18.84 | 14.45 | 8.45 | 1.32 | 15.25 |
| (proposed) | BS Laplacian | **3.53** | **3.13** | **2.66** | **2.11** | **2.68** | **19.30** | **14.74** | **8.78** | 1.55 | **15.85** |
| Oracle SIBF | TV Gaussian | 3.58 | 3.21 | 2.80 | 2.39 | 2.75 | 20.62 | 17.03 | 12.25 | 6.54 | 17.99 |
| | BS Laplacian | 3.58 | 3.21 | 2.80 | 2.39 | 2.75 | 20.45 | 17.05 | 12.33 | 6.59 | 18.00 |
| Reference generated with the DNN | | 3.14 | 2.83 | 2.43 | 1.91 | 2.61 | 18.48 | 13.89 | 8.70 | **2.34** | 13.61 |
| Microphone #5 | | 2.93 | 2.51 | 2.10 | 1.72 | 2.18 | 14.05 | 8.03 | 2.03 | -3.93 | 7.54 |
| CHiME4 SE baseline (BLSTM GEV) [1] | | | | | | 2.46 | | | | | 2.92 |
| Erdogan et.al. (BLSTM MVDR) [17] | | | | | | 2.29 | | | | | 15.12 |

Subsequently, we examined the BS Laplacian model (9) to obtain the relations between the PESQ scores and iteration times (1, 2, 5, and 10) for various reference weights ($\alpha = 10^{-2}$, 1, $10^2$, and $10^4$). Figure 4 shows a subset of the relations: BG×2.0 (left) and BG×0.25 (right) scenarios, and the scores of the reference and TV Gaussian model ($\beta = 8$ only) with dotted and broken horizontal lines, respectively.

In the BG×2.0 scenario, the PESQ scores were improved for all reference weights as the iteration time increased. Remarkably, a smaller weight resulted in a higher improvement. However, in the BG×0.25 scenario, the cases $\alpha = 1$ and $10^{-2}$ showed decreasing tendencies except for the first iteration. We also observed that the tendencies of the BG×1.0 and BG×0.5 scenarios (not shown here) were close to those of the BG×2.0 and BG×0.25, respectively.

From these results, we chose $\alpha = 10^2$ as the best parameter as it showed stably increasing tendencies.

### 4.4. PESQ and SDR evaluation

Next, we measured the signal-to-distortion ratio (SDR) and PESQ using the best parameter sets: $\beta = 8$ for the TV Gaussian model and $\alpha = 10^2$ for the BS Laplacian model with 10 iterations. To estimate the potential performance of the SIBF, we also attempted to use the magnitude spectrogram of a clean speech as the ideal reference. This was termed as the *Oracle SIBF*, whereas the use of the reference generated by the DNN was referred to as the *NN-SIBF*. To compare the SIBF with other methods, we also used the CHiME3-simulated evaluation set.

The results are shown in Table 2, which also includes scores of the reference and observation with Microphone #5. The last two rows demonstrate the scores of the CHiME4 SE baseline [1] and BLSTM-based MVDR [17], which were probably trained using a dataset similar to the one in this study. From this table, we confirmed the following:
1. The NN-SIBF outperformed the references in most cases, except for the SDR in the BG×2.0 scenario.
2. The BS Laplacian model outperformed the TV Gaussian model in the NN-SIBF, whereas both models showed almost identical performance in the Oracle SIBF.
3. For the scores in the evaluation set, the NN-SIBF outperformed the conventional DNN-based methods [1] [17], despite using the same training dataset.

## 5. Discussion

In this section, we discuss the following three aspects.

The first aspect is the effectiveness of utilizing both dependence and independence. This can be verified by the fact that the NN-SIBF outperformed the reference in most cases. This suggests that if we have a spectrogram generated by a target-enhancing method, we can obtain a more accurate target signal than it by using it as the reference in the SIBF.

The second aspect is the relationship between the accuracy of the reference and performance of the target extraction. We can conclude that the more accurate reference achieves better target extraction performance because the Oracle SIBF scores were much higher than those of the NN-SIBF in all the experiments. In other words, inaccurate references often result in limited improvements, as shown in the BG×2.0 scenario in Table 2. However, Figure 4 indicates that even with such inaccurate references, the BS Laplacian model can still improve performance after multiple iterations if the proper reference weight (e.g., $\alpha = 10^{-2}$) is chosen. This figure also suggests that the balance between dependence and independence should vary according to the accuracy of the reference, although this remains an open issue.

The third aspect is how to further improve the target extraction performance. From the above discussion, there are at least two options. One involves improving the accuracy of the reference using state-of-the-art speech enhancement methods to generate a reference. The other involves improving the source models by refining the proposed models, such as automatic parameter tuning, as well as examining another source model type.

## 6. Conclusions

In this study, we proposed a novel method for target signal extraction, i.e., the SIBF. The method used a rough magnitude spectrogram of the target signal as the reference to extract the target signal more accurately. For this extraction, we extended the framework of the deflationary ICA by considering the similarity between the reference and extracted target signal, as well as the mutual independence of the potential sources. To solve the extraction problem by maximum-likelihood estimation, we developed two source model types that could reflect the similarity: the TV Gaussian and BS Laplacian models. Further, we derived corresponding rules for the extraction filters.

The advantage of the SIBF lies in its ability to extract a more accurate target signal than the spectrogram generated by target-enhancing methods, such as DNN-based speech enhancement. We verified this through experiments using the CHiME3 dataset.

Finally, the SIBF is based on the principles of the ICA, whereas it works as a beamformer. Therefore, we expect that the SIBF will further promote future research in both the DNN-based beamformer and ICA-based BSS fields.